\documentclass[a4paper,10pt]{article}
\pdfoutput=1
\usepackage{amsmath,amsfonts,amssymb}
\usepackage{calc}
\usepackage{graphicx}
\usepackage{verbatim}
\usepackage[hyperindex,breaklinks]{hyperref}
\usepackage{url}
\usepackage[width=372pt,height=588pt,top=126pt]{geometry}

\begin{document}

\begin{flushright}
ITP-UU-09/59\\
  SPIN-09/49
\end{flushright}

\begin{center}
\vspace{60pt} {\Large \bf Coupling a Point-Like Mass to Quantum
Gravity \\ \vspace{10pt} with Causal Dynamical Triangulations}

\vspace{50pt}

{\sl I. Khavkine}, {\sl R. Loll} and {\sl P. Reska}
\footnote{email: i.khavkine@uu.nl, r.loll@uu.nl, p.m.reska@uu.nl}

\vspace{24pt}

Spinoza Institute and Institute for Theoretical Physics,\\
Utrecht University,\\
Leuvenlaan 4, NL-3584 CE Utrecht, The Netherlands.\\

\vspace{48pt}

\end{center}

\begin{abstract}
\noindent We present a possibility of coupling a point-like,
non-singular, mass distribution to four-dimensional quantum
gravity in the nonperturbative setting of Causal Dynamical
Triangulations (CDT). In order to provide a point of comparison
for the classical limit of the matter-coupled CDT model, we derive
the spatial volume profile of the Euclidean
Schwarzschild-de~Sitter space glued to an interior matter
solution. The volume profile is calculated with respect to a
specific proper-time foliation matching the global time slicing
present in CDT. It deviates in a characteristic manner from that
of the pure-gravity model. The appearance of coordinate caustics
and the compactness of the mass distribution in lattice units put
an upper bound on the total mass for which these calculations are
expected to be valid. We also discuss some of the implementation
details for numerically measuring the expectation value of the
volume profiles in the framework of CDT when coupled appropriately
to the matter source.
\end{abstract}

\vspace{12pt}

\newpage

\section{Introduction}
There has been growing interest in a nonperturbative formulation
of quantum gravity in recent decades. Several candidate theories
have emerged, among which is the Causal Dynamical Triangulations
(CDT) programme. This approach implements a nonperturbative
path-integral quantization of gravity, where each contributing
spacetime history carries a well-defined causal structure. In one
of the phases of the underlying statistical model of `random
geometry' one has observed the formation of an extended universe
with good classical properties. More specifically, it has been
shown that both its Hausdorff \cite{emergence4d} and spectral
dimension \cite{SpectralDim} are four on large scales.
Furthermore, the large-scale shape of this dynamically generated
background geometry matches to great accuracy that of a de Sitter
universe \cite{PlanckianBirth}, corresponding to a universe with a
positive (renormalized) cosmological constant, and the quantum
fluctuations around it agree with predictions from a
mini-superspace model \cite{BirthPRD}.

As a next step towards making the model more realistic, we want to
study matter coupling using CDT. In the framework of dynamical
triangulations, it is straightforward to set up dynamical, coupled
gravity-matter systems by extending the sum over all geometries to
a double-sum over all geometrical \emph{and} matter field
configurations (for a spin, scalar or gauge field, say). This has
already been demonstrated in the corresponding \emph{Euclidean}
quantum gravity models in four dimensions
\cite{euclideanmatter,gaugematter,z2matter,abeliangauge,edtHorata}.
The situation we will be studying presently is slightly different,
and assumes that a particular matter configuration has already
been arrived at, namely, one that can be approximated by a compact
mass distribution that we shall refer to as a \emph{point-like mass}.
Compared with the matter-free case, this introduces an
inhomogeneity in the geometry of the spatial slices, but preserves
spherical symmetry. Such a situation is of great physical
relevance because it corresponds to the gravitational field of any
spherically symmetric mass distribution in a universe with a
positive cosmological constant.

The relevant classical solution to the Einstein equations outside
the source is given by the Schwarzschild-de~Sitter (SdS) metric.
In this paper, we address the question of how to detect the
presence---on sufficiently large scales---of this particular
background geometry in CDT quantum gravity coupled to a point-like
mass. This implies finding a quantum observable that is well
defined in the nonperturbative, background-independent setting of
the full path integral, is sensitive to the presence of the mass,
and is potentially measurable in the computer simulations. The
dynamics of CDT quantum gravity is defined directly on the space
of geometries (in a continuum language: the space of metrics
modulo diffeomorphisms) and forces one to tackle the issue of
observables head-on, by giving measuring prescriptions for
geometric observables whose expectation values with respect to the
ensemble average over all spacetime geometries give nontrivial
results. Examples of this are the dynamical dimensions mentioned
above.

In the CDT setting, another class of geometric quantities is
accessible relatively easily, namely, those referring to the
proper-time slicing that comes with the formulation.\footnote{One
should keep in mind that geometric quantities associated with a
spatial slice in the simulations will in general not correspond
directly to properties of classical three-geometry; for this, they
will usually need to be smeared out (coarse-grained) over some
finite time-extension, cf.\ \cite{CDT1}. However, the spatial
three-volume considered below is a sufficiently robust quantity,
for which this turns out not to be necessary.} This has been used
previously in the pure-gravity theory to study both the
\emph{volume profile} (the development $V_3(\tau)$ of the spatial
three-volume as function of proper time $\tau$) and the correlator
of quantum fluctuations in the three-volume around the dynamically
generated de Sitter background spacetime
\cite{PlanckianBirth,BirthPRD}. As we shall see, the volume
profile is a geometric quantity that is modified by the presence
of the mass. We shall focus in this work on the derivation in the
continuum of the volume profile of the Euclidean Schwarzschild-de
Sitter (ESdS) solution in a proper-time slicing, which may be
compared to the values obtained from CDT simulation. To our
knowledge, no background-independent approach to quantum gravity
has so far succeeded in generating a Schwarzschild-de Sitter
geometry in the classical limit. The volume profiles we derive can
be used as a criterion to identify this spacetime in any
background-independent approach to quantum gravity in four
spacetime dimensions, when using a suitable proper-time slicing.

Unlike the previously mentioned work on dynamical matter fields,
very little has been done on coupling point or point-like masses
to quantum gravity in four dimensions. We are not aware of any
nonperturbative, background-independent approach to quantum
gravity that has been \emph{ab initio} coupled to point masses in
four dimensions. Inclusion of point particles is common in
discussions of both classical and quantum three-dimensional
gravity~\cite{Carlip2+1}. The three-dimensional theory has no
local degrees of freedom and therefore is significantly
different from the four-dimensional one, already at the classical
level. The inclusion of point particles is simplified by the fact
that their presence only creates conical
defects~\cite{Staruszkiewicz,DeserJackiw'tHooft}, and not stronger
curvature singularities like in higher dimensions. Quantization of
three-dimensional gravity with point particles has been discussed
early on in~\cite{'tHooft, MatschullWelling} and more recently in
the contexts of Loop Quantum Gravity~\cite{NouiPerez} and the
Ponzano-Regge Spin Foam model~\cite{FreidelLouapre}. In four
dimensions, a spherically symmetric quantum spacetime has been
studied in a series of papers by Husain and
Winkler~\cite{HusainWinkler1, HusainWinkler2,HusainWinkler3},
where they applied canonical quantization to a symmetry-reduced
midi-superspace model of black hole collapse from a scalar field
coupled to gravity. The Euclidean Schwarzschild-de Sitter space,
which we discuss extensively below, has been studied in the
context of black hole thermodynamics~\cite{GomberoffTeitelboim}
and the stability of de~Sitter space~\cite{GinspargPerry, Bousso}.
Schwarzschild-de~Sitter space shares many of its global properties
with de~Sitter space. An overview of the latter can be found
in~\cite{Spradlin}.

In this work, we propose to perform CDT simulations with a mass
line representing the point-like source and measure the average
volume profile to test whether the classical limit of CDT
coincides with the Euclidean Schwarzschild-de~Sitter geometry. We
derive the expected deviation from the Euclidean de Sitter profile
for pure gravity by performing a proper-time slicing of the
continuum ESdS geometry glued to an interior matter region. The
particular Gaussian normal coordinate system we work with in the
continuum, in order to mimic the CDT set-up, exhibits caustics in
the vicinity of the mass, which cannot be eliminated by extending
the coordinates to the matter region. For the numerical derivation
of the volume profiles, we neglect this problematic region by
cutting out a tube from the Euclidean Schwarzschild-de~Sitter
space whose spatial slices are balls of constant Schwarzschild
radius. This procedure gives rise to an upper bound for the mass
under consideration. The profiles for masses below this bound are
the ones that should be compared with those coming from CDT
simulations to test for the possible presence of an ESdS ground
state. A mass propagating in triangulated spacetime can be
represented by a timelike mass line on the lattice dual to the
triangulation. Our analysis reveals that in order to compare the
results derived here with those coming from simulations, the mass
line should be implemented so as to form a contractible loop on
$S^4$ (or a suitable analogue on the $S^3\times S^1$-topology used
in simulations).

The paper is organized as follows. In section~\ref{sec:CDT} we review CDT,
emphasizing its hypersurface structure and its classical limit and
describe the possibility to couple a point-like mass. Section~\ref{sec:esds}
deals with the properties of ESdS space. Starting from the metric
in static form we construct Gaussian normal coordinates to obtain
a particularly simple proper-time form of the metric. In section~\ref{sec:vols}
we derive the bound on masses accessible in simulations and derive
the volume profiles of the ESdS geometry in the proper-time
slicing for masses below this bound. Section~\ref{sec:concl} contains the
conclusion and the outlook. Appendix~\ref{sec:geodesics} contains the technical
details of the derivation of the metric in proper-time form. In
Appendix~\ref{sec:interior} we discuss the matching of an interior matter solution
to the exterior vacuum solution.

\section{Causal Dynamical Triangulations}\label{sec:CDT}

\subsection{Regularization of the gravitational path integral}

For readers unfamiliar with the quantization programme of Causal
Dynamical Triangulations, let us briefly review its motivation,
implementation and the main results it has produced to date (for
more in-depth reviews, see
\cite{Ambjorn:2005jj,Ambjorn:2006jf,Ambjorn:2009ts,Ambjorn:2010rx,Loll:2007rv}).

Building on insights from general relativity and (canonical)
quantum gravity, this approach uses nothing but standard
quantum-field theoretic principles and methods, adapted to the
situation where geometry is no longer part of a fixed background
structure, but is itself dynamical. The basic quantum-dynamical
principle it implements is the Feynman path integral, the
``superposition of gravitational amplitudes" or ``sum over
histories"
\begin{equation}
Z=\int\limits_{[g]\in\cal G} {\cal D}g\ {\rm e}^{iS^{\rm EH}[g]},
\;\;\; {\rm with}\;\;\; S^{\rm EH}=\frac{1}{G_N}\int d^4x
\sqrt{\det g} (R-2 \Lambda), \label{gravint}
\end{equation}
where each history is a spacetime geometry (a diffeomorphism
equivalence class $[g]$ of metrics on a fixed manifold $M$, with
$\cal G$ the space of all such equivalence classes), weighted with
the exponential of $i$ times its Einstein-Hilbert action. Because
of the nonrenormalizability of gravity as a perturbative quantum
field theory on a Min\-kow\-ski\-an background, such a path integral
necessarily has to be nonperturbative, which means that it must
include spacetime configurations ``far" from any classical
solution. The evaluation of the ensuing, highly non-Gaussian path
integral is technically challenging, and in the CDT approach is
addressed by using powerful lattice methods, borrowed from the
nonperturbative treatment of QCD. Adapting them to gravity implies
that the rigid lattices of gauge theory become themselves
dynamical, and actually take the form of {\it dynamical
triangulations}, because of the way the infinitely many
geometric/curvature degrees of freedom of the theory are
regularized.

Namely, the gravitational path integral is regularized by summing
over a class of piecewise flat four-manifolds, which can be
thought of as being assembled from (two types of) four-dimensional
simplices, which are simply triangular building blocks cut out of
Minkowski space. They are individually flat, but can pick up
nontrivial deficit angles after being glued together pairwise
along three-dimensional subsimplices (tetrahedra), with curvature
concentrated at two-dimensional subsimplices (triangles) where
tetrahedra meet. This does not imply that spacetime is conjectured
to consist of (Planck-sized) triangular building blocks. On the
contrary, the edge length of the simplices serves as a
short-distance cut-off and we are only interested in the {\it
universal} properties of the model as this cut-off is sent to
zero.

When evaluating the path integral, these simplicial geometries are
taken to be Euclidean, like those that are summed over in the
Euclidean gravitational path integral in its standard
definition~\cite{EQG}.
In contrast to previous Euclidean quantum gravity work,
the triangulations used in CDT have
a preferred, discrete notion of time $\tau=1,2,3,\dots $ inherited
from a class of triangulated Lorentzian piecewise flat spacetimes
by explicit Wick rotation \cite{CDT4,CDT2}. In the Lorentzian
regime, the triangulations are restricted to those consisting of a
sequence of slices with (discrete) proper-time thickness
$\Delta\tau =1$ and fixed topology. Links that lie in a spatial
hypersurface of constant integer $\tau$ are spacelike and links
connecting two adjacent spatial slices of this kind are timelike.
The restriction on the path integral histories is motivated by the
desire to eliminate the causality-violating `baby universes' in
the time direction produced in \emph{Euclidean} dynamical
triangulations~\cite{EDT}, which lead to an incorrect classical
limit because of the absence of an extended four-dimensional
geometry on large scales. After discretization and Wick rotation,
the path integral
becomes a statistical sum with Boltzmann
weights using the Regge action~\cite{Regge}, which is the
discretized version of the Einstein-Hilbert action. The first
major result of the CDT formulation was to show in exactly
solvable two-dimensional quantum gravity that the signature, i.e.,\ 
sum over Lorentzian as opposed to Euclidean geometries in the
path integral, leads to genuinely different properties of the
model (different intrinsic Hausdorff dimension, for instance)
\cite{CDT3}.

In dimension four, the regularized path integral can no longer be
evaluated by exact methods, but Monte Carlo methods must be used
to explore its continuum limit. This has led to a number of
unexpected and new results. Since the curvature is allowed to
fluctuate strongly on short scales, and since a nontrivial
limiting process is involved, it turns out that the dimensionality
of the ``quantum geometry" generated by the path integral is not
necessarily four. Only when the summed triangulations have
the causal structure described above, and the (bare) coupling
constants are chosen appropriately, does a four-dimensional
universe emerge from the quantum theory \cite{emergence4d,CDT1}.
This is the first instance in which a classical-looking universe
has been obtained from first principles within a nonperturbative
formulation of quantum gravity. Moreover, as already mentioned in
the introduction, this dynamically generated universe
macroscopically resembles a de Sitter universe, with matching
quantum fluctuations \cite{PlanckianBirth,BirthPRD}.

\subsection{Time slicing and classical limit of CDT}

The classical limit of the CDT model is considered good if, when
the length scale is large and quantum fluctuations are small, the
continuum limit of the regularized path integral reproduces the
observable predictions of classical general relativity. In order
to compare the two, one needs to phrase their respective results
in a common language, that of geometric \emph{observables}. These
are generally hard to come by, but in the case of CDT quantum
gravity there is an extra structure that comes to our help. Each
sample CDT geometry in the regularized path integral carries a
discrete time label. Since this labelling is respected by the
quantum superposition, a (possibly rescaled) version of the
discrete time parameter labelling the slices is still available in
the continuum. The reason for calling this a \emph{proper} time
comes from the fact that (i) at the discretized level, inside each
flat four-simplex one can introduce a proper-time slicing (with
respect to the Minkowskian metric of the simplex) in a way that
after gluing all of them together, the triangulated ``sandwich
geometry" between integer times $\tau_0$ and $\tau_0+1$ can be
foliated into hypersurfaces $\tau =const$, $\tau_0\leq \tau \leq
\tau_0+1$ \cite{BH}, (ii) in the continuum limit, the volume
profile of the extended universe emerging in the `well-behaved'
phase of CDT quantum gravity matches that of a continuum Euclidean
de~Sitter space as a function of cosmological proper time, if the
bare $\tau$ of the regularized geometries is rescaled by a finite
constant. More precisely, the expectation value of the volume
profile behaves to a very good approximation as
\begin{equation} \label{EuclVolProfile}
\langle V_3 (\tau) \rangle = A \cos^3 \left(\tau/B \right),
\end{equation}
for some constants $A$ and $B$ depending on the bare coupling
constants and geometric parameters of the triangulation (see
\cite{BirthPRD} for further details). The volume profile of
Euclidean de~Sitter space has exactly the same shape [cf.\
equation~\eqref{explEdSprofile} below].

The latter is of course a highly nontrivial result, however, one
needs to keep in mind that the role of $\tau$ as (constant
multiple of) proper time---in the way this notion is used in the
classical continuum theory---emerges unambiguously from CDT only
on sufficiently large scales and in the sense of a quantum average
(of a particular quantum observable). In order to understand
better the relation with the continuum situation, recall that in
the classical theory geometries with a time foliation are
naturally described in the ADM formalism~\cite{MTW}. Labelling the
spatial slices of the foliation as hypersurfaces of constant time
$t$, and choosing coordinates $x^i$ on each of them, the geometry
is specified by writing the metric in the ADM form,
\begin{equation}
ds^2 = -N(t,x)^2 dt^2 + h_{ij}(t,x)(dx^i + N^i(t,x) dt)(dx^j + N^j(t,x) dt),
\label{admmetric}
\end{equation}
with lapse function $N(t,x)$, shift vector $N^i(t,x)$ and spatial
metric $h_{ij}(t,x)$. The volume profile of the spacetime with
respect to this foliation is then given by
\begin{equation}\label{v3-def}
    V_3(t) = \int d^3x\,\sqrt{\det h}.
\end{equation}
A metric in proper-time form is one where $N=const$.
Requiring in
addition the shift vector $N^i$ to vanish, so that there are no cross
terms $dx^i dt$ and the gauge is essentially fixed, one obtains a
metric in \emph{proper-time gauge}~\cite{Teitel1,Teitel2}. The
associated coordinates are the same as Gaussian normal
coordinates~\cite{Wald,MTW} with respect to any of the spatial
hypersurfaces. Although such coordinate systems can always be set
up in the neighbourhood of
a hypersurface, they rarely
exist globally because of the formation of caustics.\footnote{Even
in flat Minkowski space, by choosing an initial hypersurface $\tau
= 0$ with typical extrinsic curvature $ K$, caustics will form
after a typical evolution time $\tau\sim K^{-1/2}$. For instance,
a sphere of radius $R$ has $K\sim 1/R^2$, while Gaussian normal
coordinates, extended to the interior, become singular at its
centre.} Therefore, taking a path integral only over those smooth
metrics which globally can be put into proper-time gauge would
appear far too restrictive.

However, this is not what is done in CDT quantum gravity, assuming
we identify the time $t$ in (\ref{admmetric}) with CDT's
$\tau$-parameter. Firstly, the ADM-decomposition (\ref{admmetric})
for differentiable, metric manifolds cannot in general be extended
beyond a single four-simplex
in piecewise flat simplicial geometries;
they are neither smooth nor
differentiable.
(The same holds for individual CDT three-slices.)
Secondly, when one follows the geodesics of
freely falling, initially hypersurface-orthogonal observers in CDT
(which are still well defined in open neighbourhoods not containing
curvature singularities), one finds that they generically form
caustics within a single time step $\Delta\tau=1$. From this point
of view, CDT histories are indeed full of caustics (and curvature
singularities), whose density only increases as the lattice
cut-off is taken to zero. Since individual path integral histories
are \emph{not} physical, this in no way contradicts the
possibility that their \emph{nonperturbative superposition} can be
a quantum geometry which on large scales behaves classically. As
we have seen for the case of Euclidean de Sitter space (EdS), the
ground state of the empty universe emerging from CDT, it is also
no obstacle to the existence of a well-defined global description
in proper-time gauge. The fact that Euclidean de Sitter space
possesses a global proper-time form which moreover has a direct
Lorentzian interpretation under the straightforward substitution
$\tau\rightarrow -i\tau$ is in a way a fortunate circumstance. If
we want to use CDT quantum gravity to describe different physical
situations, associated with a specific matter content and/or
boundary conditions, we would in general expect that making a link
to Lorentzian continuum physics will be (much) more difficult.

As we will see in subsequent sections, the inclusion of a
mass distribution already presents challenges of this kind.
Nevertheless, by requiring the mass distribution to be
sufficiently compact (point-like), the total mass $M$ to be
small and treating the situation as a one-parameter family
deviating from the known, pure-gravity case $M=0$, we are able to
quantify the consequences of $M\not= 0$ for the physical volume
profile.

\begin{figure}
\centering
\newsavebox\univbox%
\newsavebox\vprofbox%
\savebox\univbox{\includegraphics[angle=90,width=.5\textwidth,clip,trim=30 60 30 60]{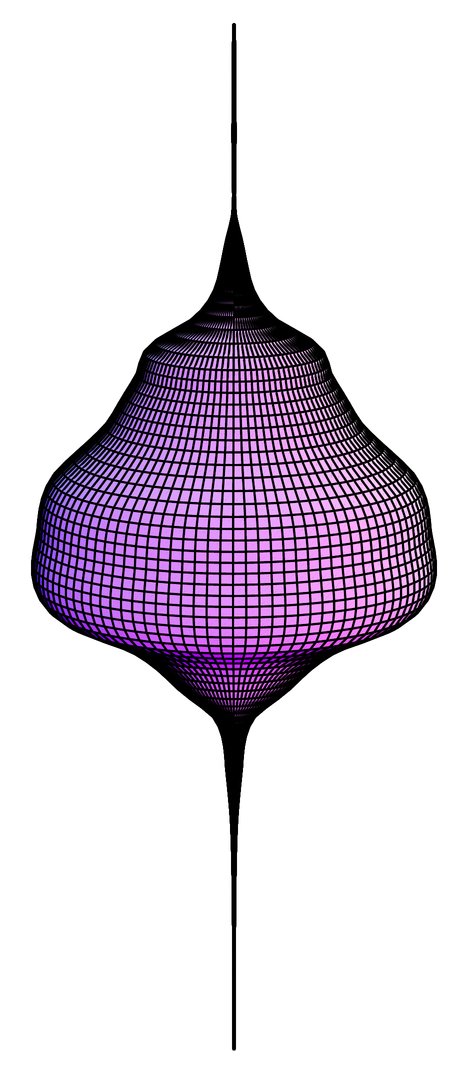}}%
\savebox\vprofbox{\includegraphics[width=.35\textwidth]{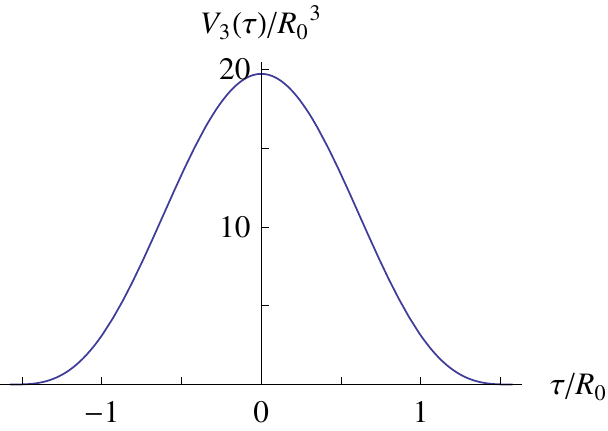}}%
(a)\raisebox{2ex-\ht\univbox}{\usebox\univbox}
(b)\raisebox{2ex-\ht\vprofbox}{\usebox\vprofbox}
\caption{\footnotesize (a) Volume profile of a typical CDT universe that
contributes to the sum over triangulations, made into a solid of
revolution by rotation about the proper-time axis. It consists of an
extended region, the `blob', and a degenerate tube.
(b) Normalized volume profile $V_3(\tau)/R_0^3$
for EdS space, cf.\ equations~\eqref{EuclVolProfile} and~\eqref{explEdSprofile},
which after a constant rescaling of $\tau$ matches that
of the expectation value $\langle V_3(\tau)\rangle$
determined in CDT simulations \cite{PlanckianBirth, BirthPRD}.
} \label{fig: CDTuniverse}
\end{figure}

Before describing the inclusion of such a point-like mass, let us
comment on spacetime topology. Computer simulations of CDT in four
dimensions are performed with compact manifolds of product
topology $I\times \Sigma^{(3)}$ or, if for simplicity the time
direction is compactified, $S^1\times \Sigma^{(3)}$. In
simulations considered so far, the spatial slices were chosen to
be topological three-spheres, $\Sigma^{(3)}=S^3$. Interestingly,
in the pure-gravity case, despite fixing the topology to
$S^1\times S^3$ at the outset, the system is driven dynamically to
a state which is as close to a four-sphere as allowed by the
kinematical constraints
(minimal, nonvanishing spatial diameter at each time step).
This is illustrated
in Fig.~\ref{fig: CDTuniverse}a, which shows the volume profile
$V_3(\tau)$ of a typical sample geometry from the regularized path
integral.
It consists of an extended universe which forms a `blob' and a
thin degenerate tube or `stalk' of minimal extension.
After subtracting the minimal stalk-volume from the data, the average
volume profile can be matched to that of EdS space (the ``round four-sphere"),
shown in Fig.~\ref{fig: CDTuniverse}b, with great accuracy.
In simulations, the period of the time
identification is much larger than the time extension of the
universe, such that this result is unaffected by the periodic boundary
conditions.

\subsection{CDT with a localized mass}

We want to generalize the above discussion by including matter.
More specifically, we are interested in the effect of a localized
mass on the ground state geometry found in Causal Dynamical
Triangulation simulations. The concept of a point mass is already problematic
in classical general relativity, in the sense of including it
consistently as a distributional source in the Einstein equations
(for a recent overview, see \cite{distribution} and references
therein). We will refrain from making this idealization and assume
an extended, but localized, spherically symmetric mass source, mostly without
discussing its internal structure. An exception is Appendix~\ref{sec:interior},
where we consider the extension of the proper-time coordinates in
the exterior to a simple interior matter distribution. In any
case, from the point of view of the regularized geometries used in
the simulations, one cannot really distinguish between a point
mass and a compact, massive object that fits inside a single
spatial simplex. In this context, we call a compact mass distribution
\emph{point-like} if its spatial extent does not exceed that of a single
triangulation simplex.

In CDT simulations, the worldline of such a mass is naturally
represented by a time-like path, transverse to the foliation, on
the dual lattice of the triangulation, as illustrated in
Fig.~\ref{fig: MassLine}.
\begin{figure} \centering
\includegraphics[scale=0.6]{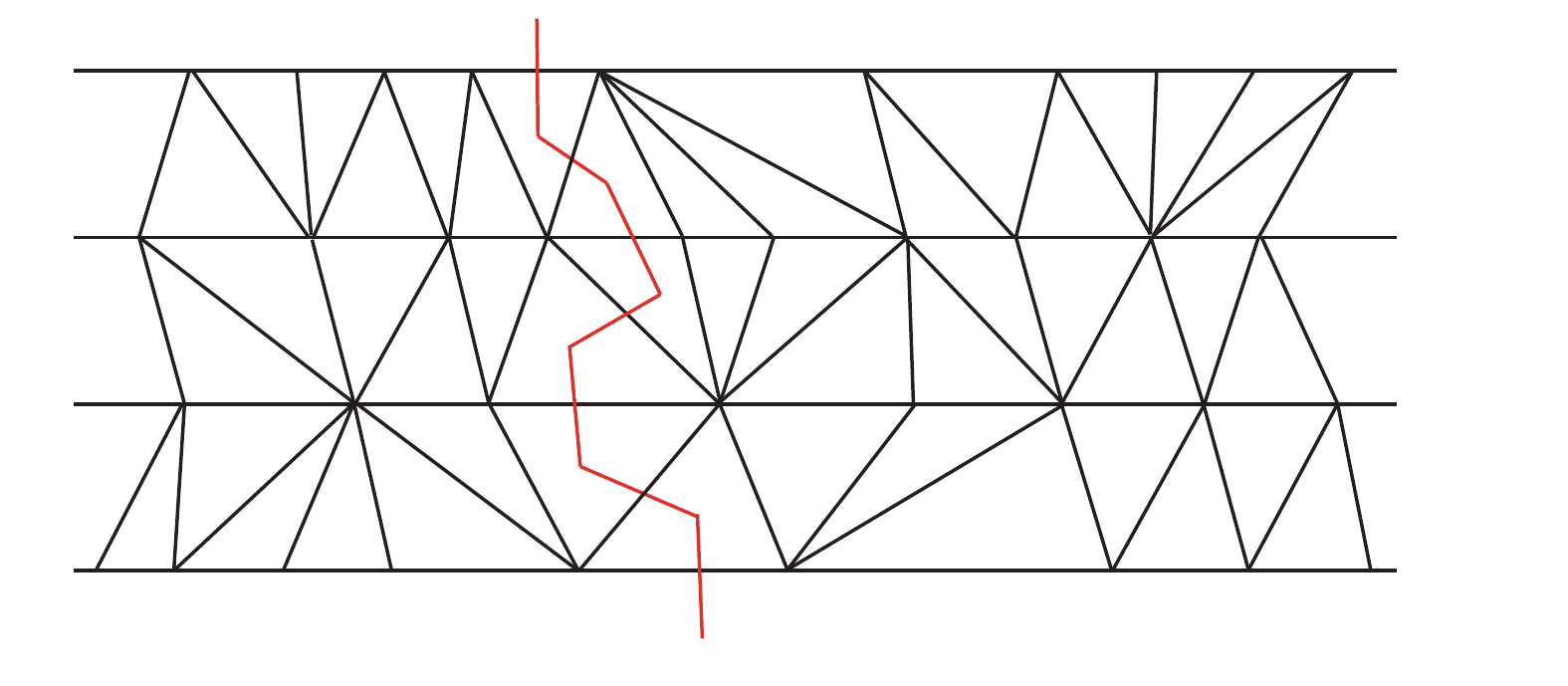}
\caption{\footnotesize Mass line on the dual lattice of a
triangulation in CDT, for simplicity demonstrated in two dimensions,
representing the worldline of a point-like mass.} \label{fig:
MassLine}
\end{figure}
The (Euclidean) action associated with this mass line is $S_p = M
L$, where $M$ is the bare mass, and $L$ is the (positive) total
length of the line in units of the lattice spacing $a$. This is
the regularized version of the continuum action associated to a
localized mass $M$, $S_p = M \int d\tau$, integrated with respect
to the proper time along its world-volume, which in turn depends
on the spacetime geometry containing the mass line. The action
$S_p$ gives an extra contribution to the Boltzmann weight of each
path-integral configuration and therefore changes the expectation
value of geometric quantities under consideration.
In particular, the volume profiles are expected to be modified.---%
The remainder of this paper is devoted to the derivation in the
continuum of the classical form of the modified volume profiles,
which could be compared to the expectation values of corresponding
quantities numerically obtained from CDT simulations.

\section{Euclidean Schwarzschild-de Sitter space}\label{sec:esds}

\subsection{Metric in static form}\label{static-form}

Written in the static form of the metric, the line element of
Euclidean Schwarz\-schild-de~Sitter (ESdS) space is
\begin{equation} \label{staticESdS}
ds^2 = f(R)dT^2 + f(R)^{-1}dR^2 + R^2d\Omega^2(\theta,\phi),
\end{equation}
where $f(R) = 1-2M/R-R^2/R_0^2$, and the coordinates $T$ and $R$
are referred to as static or Schwarzschild coordinates. The mass
of the source is $M$ and $R_0 = \sqrt{3/\Lambda}$ with $\Lambda>0$
being the cosmological constant. For a given cosmological
constant, there is an upper limit for the mass of the
Schwarzschild black hole, given by $M_N = 3^{-3/2}R_0$, the Nariai
mass~\cite{Nariai}, because the static region disappears in this
limit. We recover Euclidean de Sitter space for $M=0$, and the
Euclidean Schwarzschild metric for $\Lambda=0$. The metric has
Euclidean signature in the static region $R_+ < R < R_{++}$, where
\begin{equation}
R_+ = 6M_N  \cos\left({\frac{\alpha + 4 \pi}{3}}\right), \quad
R_{++} = 6M_N \cos\left(\frac{\alpha}{3}\right),
\end{equation}
are the locations of the black hole horizon and the cosmological
horizon, respectively, and $\alpha = \arccos\left(-M/M_N\right)$.
Also, $R_{+} \to 0$ and $R_{++} \to R_0$ as $M\to 0$.

For the Lorentzian version of the solution (\ref{staticESdS}),
the event horizon only forms if the object that
generates the gravitational field is sufficiently dense. We shall
consider a mass distribution that is glued to the exterior vacuum
region such that no event horizon is present. For now we will
focus solely on the properties of the vacuum region.

\begin{figure}
\centering
\includegraphics[width=\textwidth]{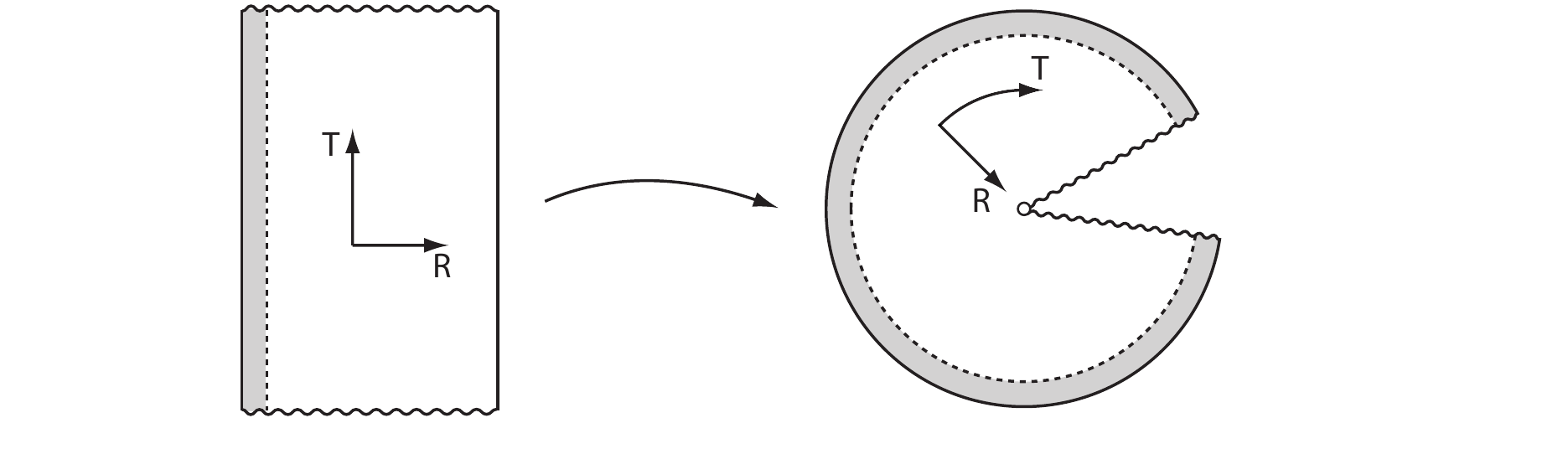}
\caption{\footnotesize Euclidean Schwarzschild-de Sitter
space can be compactified by taking
the Schwarzschild time coordinate $T$ to be periodic. The geometry
can be smoothed out in the central point $R=R_{++}$ (the location of
the cosmological horizon in Lorentzian signature) by adjusting
the periodicity of $T$. The shaded region represents the matter
region and contains no conical singularities. It is glued to the
exterior vacuum solution at a constant Schwarzschild radius
$R=R_S>R_+$, outside the horizon.
The compactified, closed geometry has
topology $S^4$.}\label{fig: TimeId}
\end{figure}

The topology of Euclidean de Sitter space is $S^4$. Suppressing
the two angular variables $\theta$ and $\phi$
in~\eqref{staticESdS} with $M=0$, we see that the two-dimensional
sheet spanned by $T$ and $0 < R < R_0$ can be wrapped
by taking $T$ to be periodic, as displayed
in Fig.~\ref{fig: TimeId} (depicting the general Euclidean
Schwarzschild-de Sitter case). The
resulting punctured disc has a potential conical singularity at
its centre, $R = R_0$, which is smoothed out by a specific choice
of the $T$-period, namely $4\pi/|f'(R_0)| = 2\pi R_0$. One obtains
the complete Euclidean manifold (for $M=0$ this is $S^4$, as we
will make explicit below) by gluing a two-sphere of radius $R_0$
to the puncture boundary.

In the case of Euclidean Schwarzschild-de Sitter space with $M\ne 0$,
the maximal Euclidean vacuum
region is spanned by $R_{+} < R < R_{++}$. Periodic
compactification along the $T$-direction introduces a different
global topology and two potential conical singularities at $R_{+}$
and $R_{++}$, which cannot be smoothed out
simultaneously~\cite{LinSoo}. Fortunately, the new topology and the
ambiguity in periodicity need not be dealt with if the manifold
consists of an ESdS exterior and an interior matter region (shaded
in Fig.~\ref{fig: TimeId}), without an inner horizon. In this
case, the periodicity is again uniquely fixed to
$T_P=4\pi/|f'(R_{++})|$, which smoothes out the geometry around
$R=R_{++}$, and the topology remains $S^4$.

As a consequence of the compactification, the mass line closes to
a contractible loop on the $S^4$. If one wants to compare with the
calculations done in this paper, this should be taken into account
when implementing the mass line in CDT simulations, rather than
using a noncontractible loop which winds around the compactified
time direction, say.

\subsection{Metric in proper-time form}\label{proper}

As has been pointed out already, the explicit derivation of the
volume profile of the Euclidean Schwarzschild-de~Sitter geometry
has to be done by adopting a proper-time gauge. In order to
achieve this form, we construct comoving or Gaussian normal
coordinates (cf.\ \cite{Wald, MTW}) from the $T=0$ surface, which
for the matter-free case coincides with half of the `equator' of $S^4$ and
will in general form a time-symmetric hypersurface. This procedure has
been previously carried out, in Lorentzian signature, for the de Sitter
case in~\cite{gautreauPRD} and for the Schwarzschild case
in~\cite{gautreauNC}.

We briefly describe the necessary calculations here, while
referring the reader to Appendix~\ref{sec:geodesics} for the
details. First, one has to integrate the radial geodesic
equations, taking initial conditions that guarantee that the
geodesics are perpendicular to and their proper time parameter
vanishes on the $T=0$ hypersurface. Then one chooses the proper
time of these radial geodesics as a new time coordinate $\tau$.
The comoving radial coordinate, $R_i$ (the subscript $i$ stands
for \emph{initial}), can be introduced as the position $R(T=0)$,
i.e. labelling each geodesic with the value of $R$ at which it
intersects the $T=0$ hypersurface. The resulting metric has
proper-time form and is diagonal,
\begin{equation} \label{LTB}
ds^2 = d\tau^2 + \frac{(\partial R/\partial R_i)^2}{f(R_i)}dR_i^2 + R(\tau,R_i)^2d\Omega^2,
\end{equation}
where the expression for the Schwarzschild coordinate
$R(\tau,R_i)$ as a function of the new proper-time coordinates is
known from equation (\ref{elliptic}).
For the Euclidean de Sitter case
we find the explicit expression
\begin{equation}
R(\tau,R_i) = R_i \cos \left(\tau/R_0\right)
\end{equation}
and the metric line element becomes
\begin{equation}\label{eds-proper}
    ds^2 = d\tau^2+R_0^2\cos^2(\tau/R_0)
        \left[\frac{dR_i^2}{R_0^2-R_i^2} + (R_i/R_0)^2d\Omega^2\right].
\end{equation}

\subsection{Domain of comoving coordinates and caustics} \label{causticsESdS}

It is well known that Gaussian normal coordinates in general fail
to cover the entire underlying manifold. Below, we determine how
much of the EdS and ESdS spaces can be covered by the coordinates
constructed in the previous section.
\begin{figure}[t]
\centering
\includegraphics[width=\textwidth]{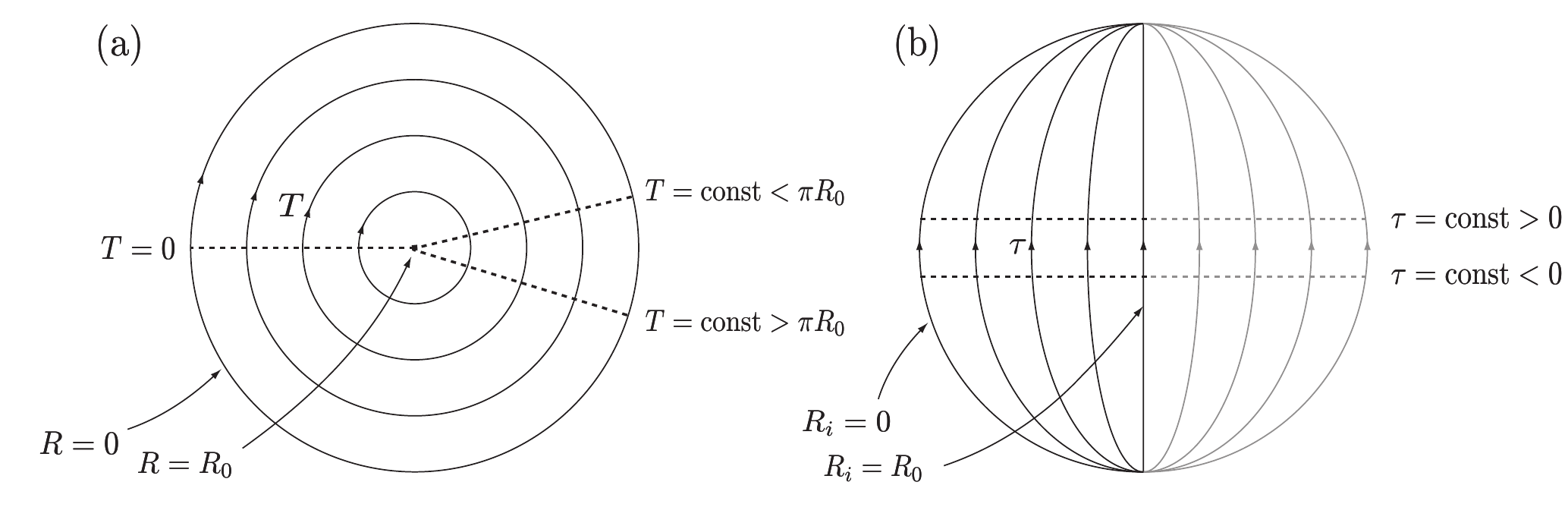}
\includegraphics[width=\textwidth]{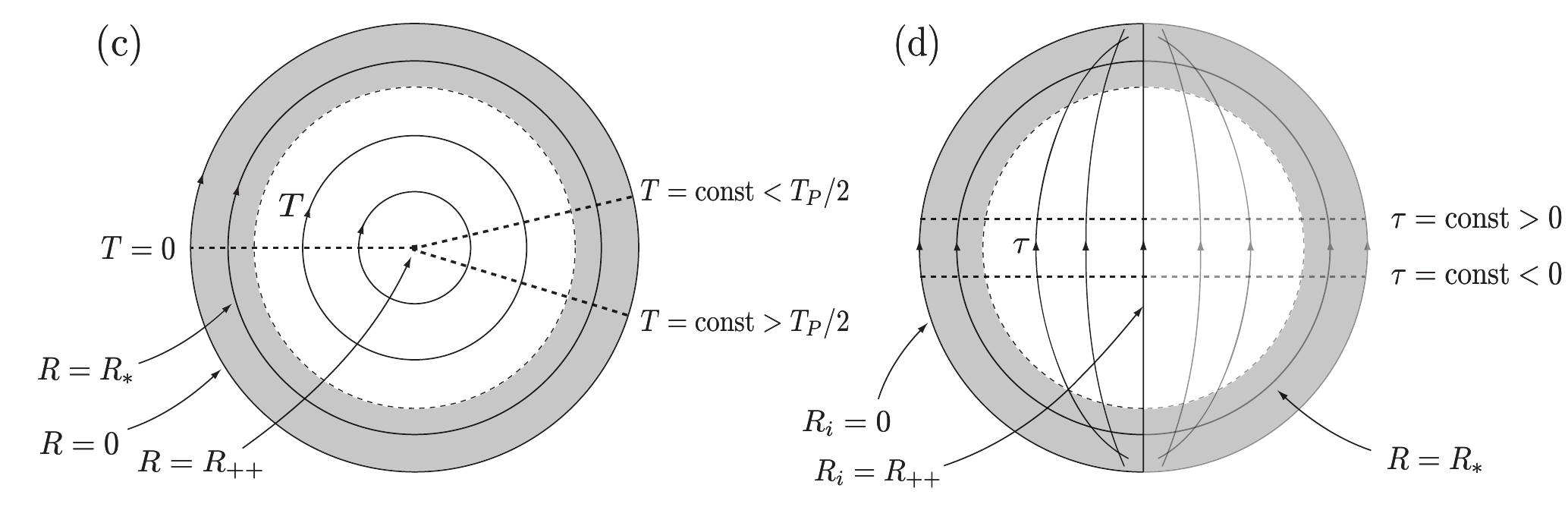}
\caption{\footnotesize Euclidean de~Sitter space (a, b) and
Euclidean Schwarzschild-de~Sitter space (c, d) (after periodically
identifying the static time $T$, with angular variables $\theta$,
$\phi$ suppressed), described in terms of (a, c) static and (b, d)
proper-time coordinates. The shaded region $R<R_S$ in (c, d) is
occupied by matter. In static coordinates, lines $R=const$ are
concentric circles, with mid point $R=R_{++}$ and outermost circle
$R=0$. In the proper-time coordinates, the lines $R_i=const$
converge at $\tau = \pm \pi R_0/2$ for (b), they start forming
caustics in the region $R<R_*$ for (d). The middle line is $R_i =
R_{++}$. Every $R_i=const$ appears once on the left half of the
disc (dark curves, perpendicular to the $T=0$ surface), and is
mirrored on the right half of the disc (light curves,
perpendicular to $T=T_P/2$).} \label{fig: Disc}
\end{figure}

First, we consider the simpler case of EdS space. The relationship
between static $(T,R)$ and comoving $(\tau,R_i)$ coordinates is
illustrated in Fig.~\ref{fig: Disc}. The disc depicted in
Fig.~\ref{fig: Disc}a is formed from $R$-$T$-space by
periodically identifying the static time $T$ such that $0\leq
T\leq 2\pi R_0$. The static coordinates $(T,R)$ cover the entire
disc. Lines of constant Schwarzschild radius $R$ are concentric
circles around the point $R=R_0$ (rather a two-sphere, with
angular coordinates taken into account), which is added to the
Euclidean manifold to make it complete. Fig.~\ref{fig: Disc}b
depicts the same disc, but now with dark vertical lines
representing constant $R_i$ (radial geodesics) and dashed
horizontal lines representing constant $\tau$. Note that only half
the disc is covered by Gaussian normal coordinates emanating from
the $T=0$ surface. The other half (which is of a lighter shade in
the figure) can be covered by reflecting the $(\tau,R_i)$
coordinates about $R_i=R_0$ or, equivalently, by constructing Gaussian
normal coordinates from the $T=\pi R_0$ surface, which smoothly joins
the $T=0$ one.

Figs.~\ref{fig: Disc}c and~\ref{fig: Disc}d parallel the above discussion for
ESdS space, though the ranges of the static coordinates change to
$0\le T\le T_P$ and $0<R<R_{++}$. Again, the right half of the disc
in Fig.~\ref{fig: Disc}d can be covered by reflection of the
$(\tau,R_i)$ coordinates
or by
constructing Gaussian normal coordinates from the $T=T_P/2$
surface. As explained in section~\ref{static-form} the geometry we
are considering has no inner horizon since the exterior vacuum is
matched to a static interior matter solution (shaded region) at
$R_S>R_+$.

The comoving coordinate system breaks down at an intersection
point of two radial geodesics, because it would associate two
distinct labels $(\tau,R_i)$, $(\tau,R_i')$ with the same physical
point. In Fig.~\ref{fig: Disc}b we see that the geodesics
intersect only at the poles, $R=0$ and $\tau=\pm\pi R_0$.
In contrast,
Fig.~\ref{fig: Disc}d shows that the geodesics
intersect (form caustics) already
for
$R<R_*$.
Fig.~\ref{fig: Disc} is schematic, the corresponding calculations
are presented in Appendices~\ref{sec:geodesics}
and~\ref{sec:interior}. Therefore, we can conclude that comoving
coordinates $(\tau,R_i)$ cover
EdS space in its entirety%
    \footnote{Strictly speaking, these coordinates fail to cover some
        lower-dimensional submanifolds. 
				But, since we are only interested in computing three- and
				four-volumes, they can be safely ignored.}%
, while only a portion of ESdS space is covered, namely for
$R>R_*$, with $R_*$ the boundary of the caustic region.

Caustics appear already in the ESdS exterior vacuum, in which case
$R_*=(MR_0^2)^{\frac{1}{3}}$. Generically, caustics persist even if the
vacuum exterior is glued to an interior matter solution of
arbitrarily low density, as explicitly shown for a specific matter
model in Appendix~\ref{sec:interior}. Thus, for convenience, from
now on $R_*$ will only refer to this \emph{vacuum caustic boundary}.
To ensure that the entire vacuum region is covered by comoving
coordinates, we require that the matter-vacuum boundary satisfies
the condition $R_S > R_*$, which puts an upper bound on the
density of the mass distribution. We only consider densities below
this critical value. This bound and the associated bound on the
total mass are discussed in detail in sections~\ref{excision}
and~\ref{mass-bound}.

We should emphasize that the occurrence of caustics does not imply
any pathologies of the underlying spacetime, but is a consequence
of the choice of a particular coordinate system. In order to
calculate the volume profile, all we need is a proper-time
slicing. Neither the vanishing of the shift vector nor the
existence of a single, global coordinate system are in principle
necessary. However, our (time-symmetric) choice of the Gaussian
normal coordinates starting at $T=0$ and $T=T_P/2$ has the
advantage of being continuously connected to a globally
well-defined coordinate system in the limiting case $M=0$. That
coordinate system was used to successfully compare classical,
continuum volume profiles to those produced by CDT simulations. It
is this successful case that we are `perturbing' about.%
\footnote{It is possible that there exist even more convenient
coordinate choices for $M\ne0$, which would avoid caustics
altogether, but we have not found them.}

The global, proper-time coordinate system on de~Sitter space is
introduced by extending the range of the radial coordinate by
introducing an angle $\psi$, $0\leq \psi < \pi$, where
\begin{equation}
R_i = R_0 \sin \psi.
\end{equation}
Every $R_i=const$ line appears twice on the disc in Fig.~\ref{fig:
Disc}b. On the left
half of the disc it is represented by an angle $\psi<\pi/2$ and
on the right half by $\psi>\pi/2$, with the middle line given by
$R_i = R_0$. The line element written in terms of $\tau$ and $\psi$ becomes
\begin{equation} \label{GlobalEdS}
ds^2 = d\tau^2 + R_0^2 \cos^2 (\tau/R_0) \left(d\psi^2 + \sin^2 \psi\ d\Omega^2\right).
\end{equation}
Note that this is the line element of the round four-sphere, where
the geodesic $R_i=0$ has become the location of the coordinate
singularity $\psi=0$. Spatial sections of this four-dimensional
Euclidean geometry are three-spheres, which is consistent with CDT
simulations. From this metric, or already from the line
element~\eqref{eds-proper}, we can immediately derive the volume
profile
\begin{equation} \label{explEdSprofile}
V_3 (\tau) = 2 \pi^2 R_0^3 \cos^3 \left(\tau /R_0 \right)
\end{equation}
for EdS space, which we have referred to earlier
in equation (\ref{EuclVolProfile}). We now turn to the calculation
of the volume profile of ESdS space.

\section{Volume profiles}\label{sec:vols}

\subsection{Cutting out the vicinity of the mass-line}\label{excision}

In computer simulations, the volume profile is determined by
counting spatial simplices per slice of constant proper time for
the individual sample geometries and taking the average value over
the ensemble. For simulations with a mass line,
the average of the discrete geometries should approximate
the Euclidean Schwarzschild-de~Sitter space well away from
the mass, but poorly close to it. Thus it makes sense to discard the simplices
pierced by the mass line and to excise a corresponding thin `tube'
surrounding the mass from the four-dimensional continuum geometry. Only
the volume profiles of the remaining regions will be compared.
On every
spatial slice $\tau=const$ we choose to cut out the region inside
of a ball of a certain Schwarzschild radius $R$, whose area is
$4\pi R^2$. Of course, this choice of the radius depends on the
mass and only for small masses can we expect a good match between the
continuum and discrete picture. A more detailed analysis on the
mass bound will be presented in the following section.

An important property of this prescription is that the surface of
this region (which has topology $S^2\times S^1$ for compactified
Schwarzschild time) is invariantly defined, and mapped into itself
under the flow of the time-like Killing vector. This agrees with
the discrete picture, where the sequence of four-simplices cut out
of the four-geometry representing the mass line has the topology
of a tube. A strict implementation of the classical continuum
set-up on the simplicial lattice would allow only for mass lines
whose boundary is everywhere time-like, which would limit the
types of tube geometry that can occur in a single time step. This
would be associated with an excised region per time step of some
typical, average surface area $O(1)$ in discrete units, even
though the internal geometry of the excised region is considered
unspecified.
Other prescriptions for cutting out the matter region in
the classical continuum geometry are in principle possible.
However, they must not deviate from the Schwarzschild prescription
more than the scale set by the discretization (the lattice
spacing) which is the size of the error already inherent in the
triangulation procedure.

In view of the quantum nature of the CDT path integral, simulations
may have to include more general mass lines, which can wind around
longer in a given time step or are even allowed to run backwards
in time.

\subsection{Derivation of a mass bound}\label{mass-bound}

We emphasized that due to caustic formation in the continuum
picture it is necessary to cut out a certain region of the
Euclidean manifold. Now, it is important to carefully translate
the excised region from the continuum picture to the
discrete one in order to reliably compare the volume profiles. For
this we relate the physical parameters on both sides. In the
continuum there are only two parameters, the mass $M$ and the
cosmological constant $\Lambda$, which sets the cosmological
radius $R_0=\sqrt{3/\Lambda}$. Newton's constant $G_N$ has been
set to one.
In CDT simulations performed so far, the directly specifiable parameters%
    \footnote{Another parameter specifiable in simulations is the
    directional asymmetry parameter $\Delta$, though its value is not
    accessible in a classical geometry. For the discrete computations
    presented in this subsection, we will assume for definiteness that $\Delta =0$, such
    that the Euclideanized four-simplices are all equilateral \cite{CDT2}.} %
are Newton's constant, the mass, and the number of four-simplices
$N$. The lattice spacing $a$ is introduced to relate dimensionless
simulation parameters to dimensionful physical parameters. We
relate simulation and continuum parameters by comparing geometric
observables in the continuum limit. For instance, for Euclidean
de~Sitter space, the total four-volume is
$V_4=\frac{8\pi^2}{3}R_0^4$, from~\eqref{explEdSprofile}, while on
the triangulation side we have $V_4=\frac{\sqrt{5}}{96} a^4N$,
leading to the relationship
\begin{equation} \label{aR_0-ratio}
    a/R_0 = \left(\frac{8\cdot 96\pi^2}{3\sqrt{5}}\right)^{1/4} N^{-1/4} \sim 5.8 N^{-1/4}.
\end{equation}
Typical values of $N\sim 3\times 10^5$ give $a/R_0\sim 1/4$.

Another important relationship is between the lattice spacing $a$
and the maximum mass $M$ accessible to simulations. To find the
possible range of masses we recall that the vacuum caustic
boundary $R_*=3M^{1/3}M_N^{2/3}$ must fall within the matter
region, $R_*<R_S$.
On the other hand, in order to compare directly with the
calculations done in this paper, one should consider mass
distributions that are sufficiently compact to fit inside a region
of the same size as the simplex whose volume we neglect, that is
$R_S < a$.
The resulting condition $R_* \leq a$ can be
expressed as an upper bound on the mass $M$:
\begin{equation} \label{MassBound}
M ~~ \leq ~~ 3^{3/2} \left(\frac{a}{R_0}\right)^3 M_N
    ~~ =    ~~ \frac{1}{3}a^3\Lambda
    ~~ \sim ~~ a \sqrt{\frac{8\cdot 96\pi^2}{3\sqrt{5}N}}
    ~~ \sim ~~ 33.6\frac{a}{\sqrt{N}}.
\end{equation}
The first is the most useful form of the bound for the numerical
calculations carried out in Appendix~\ref{sec:geodesics}
and section~\ref{profiles}, where, as we shall see, the dimensionless
parameter $\epsilon=M/M_N$ determines the shape of the volume
profiles, while the last gives an estimate for the maximum
accessible mass in terms of simulation parameters, with the
approximation coming from direct use of
equation~\eqref{aR_0-ratio}. Since $a/R_0$ is expected to be
small, the deviations of the volume profile from the Euclidean de
Sitter case for accessible masses are also likely to be small.

\subsection{Derivation of the volume profiles} \label{profiles}

Having discussed the range of validity of the excision we can move
on to the derivation of the volume profile which can be used to
test the classical limit of matter-coupled CDT quantum gravity and
constitutes the main result of our work. From the line
element~\eqref{LTB} and definition~\eqref{v3-def}
we obtain an integral expression for the total
three-volume of the vacuum region,
\begin{equation} \label{FinalProfile}
V_3 (\tau) = 8\pi \int_{R_i^\mathrm{min}(\tau)}^{R_{++}} dR_i
\frac{R^2(\tau,R_i) R'(\tau,R_i) }{\sqrt{f(R_i)}},
\end{equation}
where $R'=\partial R/\partial R_i$, a factor of $4\pi$ comes from the
angular integration, and an extra factor of $2$
takes the doubling of the $R_i \leq R_{++}$ region into account,
as explained in section~\ref{causticsESdS}. The function
$R_i^\mathrm{min}(\tau)$ is the cut-off condition $R=a$ in
proper-time coordinates.

In the continuum, increasing the mass while keeping $R_0$ (and
therefore the cosmological constant $\Lambda$) fixed changes the
total four-volume. In CDT simulations, however, the four-volume
(that is, the number of four-simplices) is usually kept fixed for
technical reasons.\footnote{The path integrals for fixed
four-volume and fixed cosmological constant are related by a
Legendre transformation.} In order to facilitate comparison with
CDT results, we invert the relationship between $R_0$ and $V_4$ in
the continuum for fixed mass $M$. First, note that the function
$R$ in the definition of the volume profile (\ref{FinalProfile})
and the four parameters $\tau$, $R_i$, $M$ and $R_0$ on which it
depends all have the dimension of length. Hence, we can write
\begin{equation}
R(\tau,R_i,M,R_0) = R_0 F(\zeta, R_i/R_0,\epsilon),
\end{equation}
for some dimensionless function $F$ that depends only on the
dimensionless parameters $\zeta=\tau /R_0$, $\epsilon = M/M_N =
3^{3/2} M/R_0$ and the ratio $R_i/R_0$. The volume profile can
therefore also be written as
\begin{equation}\label{RenESdSprofile}
V_3 (\tau,\epsilon) = R_0^3 G(\tau/R_0,\epsilon),
\end{equation}
with another dimensionless function $G$ whose explicit form is to be
evaluated. Note that the lower integration limit
in~\eqref{FinalProfile} introduces an additional dependence on the
lattice length $a$ that we suppress here. From
equation~\eqref{explEdSprofile} it follows that
$G(\zeta,0)=2\pi^2\cos^3\zeta$ for the case $M=0$. In terms of the
function $G$ the four-volume is
\begin{equation} \label{HofEps}
V_4 (\epsilon )=
\int\limits_{-\tau_\mathrm{max}(\epsilon)}^{\tau_\mathrm{max}(\epsilon)}
d\tau\ V_3 (\tau,\epsilon) = R_0^4
\int\limits_{-\tau_\mathrm{max}(\epsilon)/R_0}^{\tau_\mathrm{max}(\epsilon)/R_0}
d\zeta\  G(\zeta,\epsilon) \equiv R_0^4 H(\epsilon).
\end{equation}
In these expressions the integration limit is given by the time
$\tau_\mathrm{max}(\epsilon)$ where the three-volume becomes zero.
For Euclidean de Sitter space the value of the function
$H(\epsilon)$ before cutting out the tube is $H(0)=8\pi^2/3$. If
$V_4^*$ is the fixed four-volume used in a simulation, we have to
adjust $R_0$ depending on the value of $\epsilon$ by setting
$R_0(\epsilon) = \left(V_4^*/H(\epsilon)\right)^{1/4}$.
\begin{figure}
\centering
\includegraphics[scale=0.9]{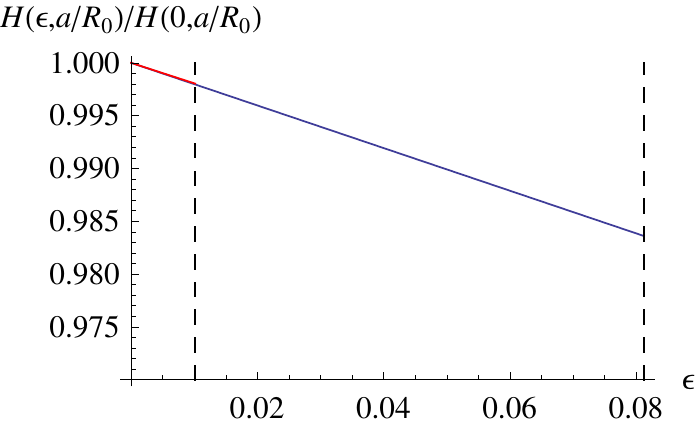}
\caption{\footnotesize Plots of the dependence of the scaled total
four-volume on the parameter $\epsilon$ for the two cut-off values
$a/R_0=1/8$ [red line, ranging from zero to the first vertical line at
$\epsilon = 3^{3/2}(1/8)^3$] and $a/R_0=1/4$ [blue line, ranging from zero to
the second vertical line at $\epsilon = 3^{3/2}(1/4)^3$]. Both
functions are normalized to one at $\epsilon=0$. In the displayed
region the curves are linear to very good approximation; the
cut-off dependence is essentially negligible.} \label{fig: HVolume}
\end{figure}%
The rescaled four-volume $H(\epsilon)$ can be easily evaluated
analytically in static coordinates:
\begin{align}\label{HofEpsExact}
H(\epsilon,a/R_0) &= V_4/R_0^4 = \frac{4\pi}{R_0^4}\int_0^{T_P} dT \int_a^{R_{++}} dR\ R^2 \\
&=\frac{8\pi^2}{3}\frac{8\cos^3(\alpha(\epsilon)/3)-3^{3/2}(a/R_0)^3}{\left|\frac{3\epsilon}{4\cos^2(\alpha(\epsilon)/3)}-6\cos(\alpha(\epsilon)/3)\right|}, \nonumber
\end{align}
where $f(R)$, $R_{++}$, $T_P$, and $\alpha(\epsilon)=\arccos(-\epsilon)$
were introduced in section~\ref{static-form}. We checked the
numerical evaluation of the volume profiles $V_3(\tau,\epsilon)$,
as described below, by comparing the value of $H(\epsilon,a/R_0)$
obtained from numerical quadrature of~\eqref{HofEps} to its exact
value in~\eqref{HofEpsExact} and found that our numerics are
reliable.

For small $a/R_0$, and over the correspondingly small range of
$\epsilon$, the total four-volume is well approximated by
\begin{equation}\label{HofEpsLin}
    \frac{H(\epsilon,a/R_0)}{H(0,a/R_0)}
        \sim 1-3^{-3/2}\left[1+3(a/R_0)^3\right]\epsilon ,
\end{equation}
as can be seen from Fig.~\ref{fig: HVolume}.

\begin{figure}
\centering \raisebox{110pt}[5pt]{(a)}
\includegraphics[scale=0.9]{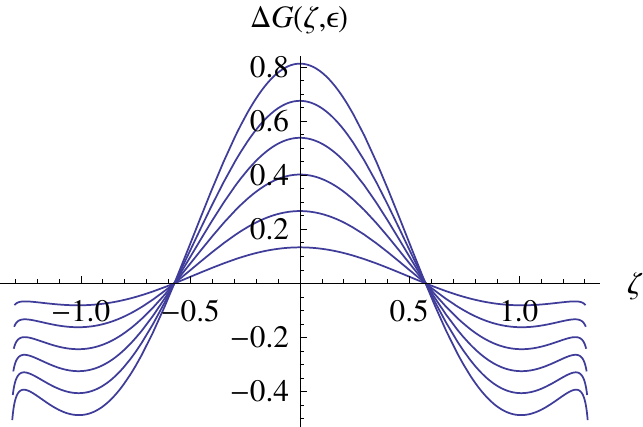}
\raisebox{110pt}[5pt]{(b)}
\includegraphics[scale=0.9]{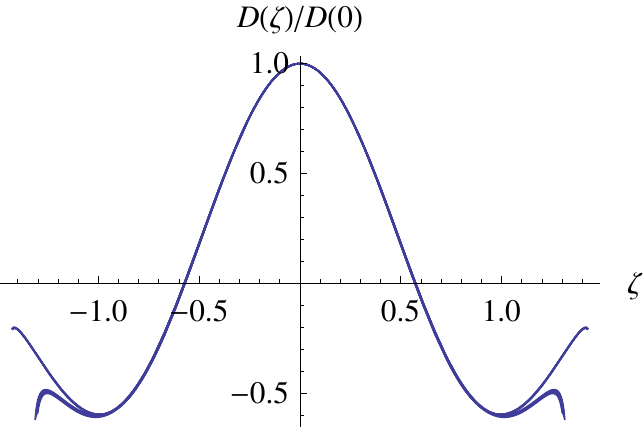}
\caption{\footnotesize (a) Difference, $\Delta
G(\zeta,\epsilon)=G(\zeta,0)-G(\zeta,\epsilon)$, between the pure
EdS volume profile rescaled to a fixed four-volume and the
rescaled volume profiles for $\epsilon =
\epsilon_\mathrm{max}/7,\ldots,6\epsilon_\mathrm{max}/7$, with
cut-off at $R=a=R_0/4$ and $\epsilon_\mathrm{max} =
3^{3/2}(a/R_0)^3$. (b) Repeating the difference measurements of
(a) for both $a/R_0=1/4$ and $a/R_0=1/8$ yields a total of 12
curves, which we have normalized by letting them all go through
the point $(0,1)$. We observe that they are multiples of a single
universal curve, modulo some weak cut-off artifacts near
$\zeta=\pm\pi/2$ that quickly disappear for smaller values of
$a/R_0$.} \label{fig: RenDeltaCutESdS}
\end{figure}%

In Fig.~\ref{fig: RenDeltaCutESdS}a we have plotted the difference
between the rescaled pure Euclidean de~Sitter profile $G(\zeta,0)$
and the rescaled profiles $G(\zeta,\epsilon)$ for different values
of $\epsilon > 0$, with the same interior region cut out, in the
mass range given by~\eqref{MassBound} and for same fixed
four-volume. They were obtained in Mathematica by first dividing
the positive region of the $\tau$-interval into 200 equidistant
parts. For each $\tau$ the integral~\eqref{FinalProfile} is
computed by numerical quadrature with the integrand given by cubic
spline interpolation over 400 equidistant points. %
		These points are obtained by solving numerically the implicit
		relation~\eqref{elliptic} for $R$ and by finite differences to find $R'$. 
Having obtained the points of the
$V_3(\tau,\epsilon)$ curve, we perform another cubic interpolation
and numerically integrate the resulting function to get
$H(\epsilon)$, according to~\eqref{HofEps}. This allows us to
rescale the profiles and obtain $G(\zeta,\epsilon)$ from
equation~\eqref{RenESdSprofile}. The curves in Fig.~\ref{fig:
RenDeltaCutESdS}a appear to scale linearly in $\epsilon$. Indeed,
after a rescaling linear in $\epsilon$, all of them collapse onto
the single curve (modulo cut-off artifacts) plotted in
Fig.~\ref{fig: RenDeltaCutESdS}b. Since we are restricted to small
values of the parameter $\epsilon$, it is not surprising that we
are in the linear regime, where rescaled profiles are well
approximated by the first two terms of the Taylor expansion in
$\epsilon$,
\begin{equation}
\label{taylor}
   G(\zeta,\epsilon)
        = 2\pi^2\cos^3\zeta - \epsilon D(\zeta) + O(\epsilon^2).
\end{equation}
In Fig.~\ref{fig: RenDeltaCutESdS}b we show $D(\zeta)/D(0)$, that
is, the linear coefficient in (\ref{taylor}), normalized such that
its value at $\zeta=0$ is given by $1$.

Numerical simulations produce volume profiles $V_3(\tau)$ in
lattice units. Rescaling the range of $\tau$ to $[-\pi/2,\pi/2]$
fixes the value of $R_0$ in lattice units and defines the rescaled
volume profile $G(\zeta)$ through equation~\eqref{RenESdSprofile}.
We believe that subtracting $G(\zeta)$ from $2\pi^2\cos^3\zeta$
and normalizing this difference to be $1$ at $\zeta=0$ should
reproduce the curve $D(\zeta)/D(0)$ plotted in Fig.~\ref{fig:
RenDeltaCutESdS}b, thereby establishing a good classical limit of
matter-coupled CDT. The corresponding value of $\epsilon$ can be
obtained from comparing the normalized simulation four-volume
$V_4^*/R_0^4$ to $H(\epsilon)$ in~\eqref{HofEpsExact}
or~\eqref{HofEpsLin}.

\section{Conclusions and Outlook}\label{sec:concl}

In any theory of quantum gravity, it is notoriously difficult to come up
with ``observables'', that is, quantities with an invariant geometric
meaning, which may eventually be related to physical observations.
Besides their obvious use in bridging between theory and phenomenology,
they play an important role at the current stage, when we are
not yet in possession of a complete, nonperturbative formulation
of quantum gravity. This role is at least two-fold. First, appropriately
coarse-grained geometric observables can provide nontrivial tests of whether
a proposed nonperturbative theory possesses a well-defined classical
limit, and whether in this limit it
reproduces the physics of classical general relativity correctly. Second,
evaluating an observable which explicitly probes the quantum regime of
the theory can be a means of comparing different candidate
theories of quantum gravity.

A prominent example of both of these uses is the so-called
spectral dimension of spacetime, measured on short and large
scales. Its expectation value was first studied in Causal
Dynamical Triangulations \cite{SpectralDim,CDT1}, exhibiting a
characteristic scale dependence. On large scales, the expected
classical value of four is reproduced, which decreases to
two\footnote{More precisely, a value compatible with two, taking
into account the error bars of the Monte Carlo simulations.} when
approaching the Planck scale, indicating strong deviations from
classicality. This is a highly nontrivial result which has since
been reproduced in at least two completely different formulations
of quantum gravity \cite{laureu,horava}, stimulating further
research into a common origin of this seemingly universal
behaviour \cite{carlip,cdthor}.

This example illustrates how observables of this type can yield
valuable information about the quantum theory. Unfortunately, they
are rather rare. In the present work, we have looked at another
geometric quantity which has been studied previously in CDT, the
three-volume profile, which makes explicit use of the proper-time
foliation. In a first attempt to try and quantify the effects of
matter on geometry in this framework, we have analyzed how the
volume profile can be expected to change under insertion of a
point-like mass, as a function of the particle mass $M$, if the
corresponding ground state geometry which is generated dynamically
by CDT is related to the Schwarzschild-de Sitter geometry.

As we have seen, the analysis involved several nontrivial steps,
despite the relatively simple and static nature of the classical
metric. The difficulties have to do with the nonlocal nature of
the volume profile, which requires a careful treatment of boundary
conditions and regions of validity of the coordinate systems one
has to use in a continuum calculation. Another difficulty derives
from having to define a quantity which is geometric, i.e.,\
independent of any particular coordinate choice, to be able to
compare with the (coordinate-free) set-up of the CDT simulations.
Both of these issues are characteristic for quantum observables in
gravity.

Our specific construction involved the use of a system of Gaussian
normal coordinates. On the one hand, this gave us a relatively
good control on some of the global properties of the Euclidean
Schwarzschild-de Sitter space like at the (Euclideanized)
cosmological horizon, on the other hand the coordinates do not
cover the complete region outside the source. However, when
restricting the mass to be small, the proper-time foliation is
nearly global in the sense that neglecting the contribution to the
spatial volume from the region inside of a certain small
Schwarzschild radius, we obtain an approximation to the actual
volume profiles while preserving their characteristic deviation
from the Euclidean de Sitter profile. The approximation is
equivalent to neglecting those simplices that contain the
mass-line when determining the average spatial volume in the
computer simulations. Let us point out that the deviations we have
computed are small and one will need good control of the numerical
errors to measure them. We also found that the correct way of
implementing the mass line in the simulations, if one wants to
compare to our calculation, is by representing it by a closed
contractible loop on the geometry which has a four-sphere
topology.

We regard the present work as a step towards understanding the dynamics of
coupled systems of matter and geometry in nonperturbative quantum gravity,
about which there is currently little known, since most candidate theories
have focussed their efforts on the pure-gravity situation. As we have already
mentioned above, it is possible that our treatment can be improved, to cover
a larger region of spacetime and/or the case of larger masses.
In addition, it would be interesting to derive the volume fluctuations
from a mini-superspace action in the same way as has been done for Euclidean de Sitter
space \cite{BirthPRD} and check whether the agreement between the analytical and numerical
calculations persists.

\vspace{.7cm}

\noindent {\bf Acknowledgements.} We thank T.~Budd, S.~Butt and
A.~G\"{o}rlich for discussion.---R.L.\ and P.R.\ were partially
supported through the European Network on Random Geometry ENRAGE,
contract MRTN-CT-2004-005616. R.L.\ acknowledges support by the
Netherlands Organisation for Scientific Research (NWO) under their
VICI program. I.K.\ was supported by the National Science and
Engineering Research Council of Canada (NSERC).

\appendix

\section{Geodesics in ESdS space}\label{sec:geodesics}

\begin{figure}
\centering
\includegraphics[scale=0.9]{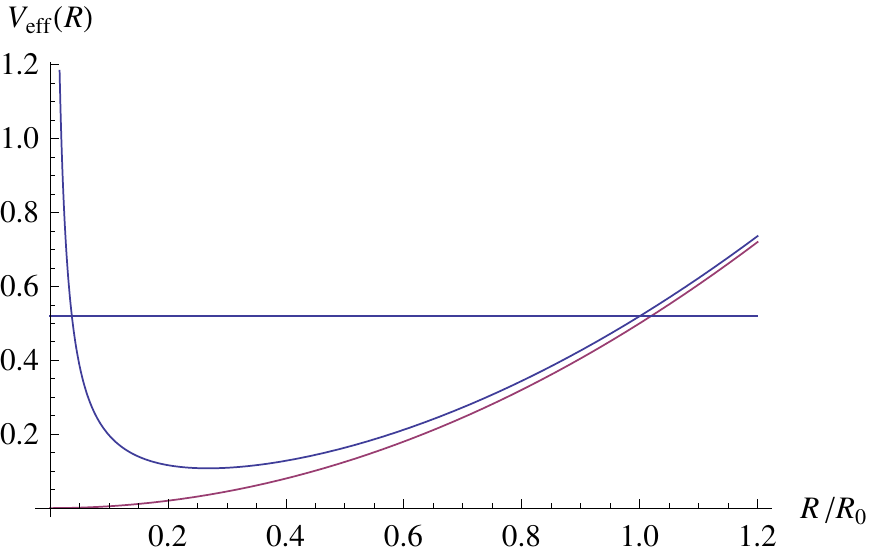}
\caption{\footnotesize The effective potential for radial geodesic
motion in ESdS space for $M = M_N/10$ (upper curve with horizontal
line to illustrate turning points for the maximum total energy)
and for $M = 0$ (lower curve). These potentials determine the
evolution of the $R_i = const$ geodesics which form the
proper-time coordinate system.} \label{fig: Veff}
\end{figure}

In this appendix we derive the line element on Euclidean
Schwarzschild-de~Sitter space in terms of Gaussian normal
coordinates, Eq. (\ref{LTB}), starting from the static form, Eq.
(\ref{staticESdS}). For this we first determine the radial
geodesic equations for ESdS space.\footnote{The complete analytic
solution for geodesics in Schwarzschild-de Sitter space was
constructed in \cite{HaLa1,HaLa2}.}
The Killing vector $\xi=\partial/\partial T$ yields a conserved
quantity along the geodesics,
\begin{equation}
E = g_{\mu\nu} \xi^\mu \frac{dx^\nu}{d\tau} = f(R)\dot{T},
\end{equation}
where the dot refers to differentiation with respect to $\tau$. We
will refer to $E$ as the energy parameter. The geodesic equations
are then
\begin{equation}
\frac{dR}{d\tau} = m\sqrt{f(R)-E^2}, \qquad \frac{dT}{d\tau} =
\frac{E}{f(R)},
\end{equation}
where the $m=\pm 1$ distinguishes motion in Euclidean proper-time
$\tau$ with increasing and decreasing $R$. Combining the two
geodesic equations we find the proper time element along each
geodesic,
\begin{equation}
d\tau = EdT + m \frac{\sqrt{f(R)-E^2}}{f(R)}dR.
\end{equation}
While $E$ is constant on a given geodesic, it may assume different
values on different geodesics. For now, we take it as a yet to be
specified function of coordinates $E(T,R)$. If we consider the
two-dimensional $R$-$T$-plane to be foliated by non-intersecting
geodesics, we can eliminate the $T$-coordinate in favour of their
proper time $\tau$ and obtain the metric~\cite{MartelPoisson}
\begin{equation} \label{PGmetric}
ds^2 = d\tau^2 + \frac{1}{E(T(\tau,R),R)^2} \left(dR -
m\sqrt{f(R)-E(T(\tau,R),R)^2}d\tau \right)^2 + R^2d\Omega^2.
\end{equation}
Note that the metric is already in proper-time form. To set the
shift vector to zero, and obtain a Gaussian normal coordinate
system, we replace the $R$-coordinate by a comoving radial
coordinate. We introduce the comoving radial coordinate by first
observing that radial geodesic motion of a test body in ESdS space
corresponds to motion in the effective potential
\begin{equation}\label{VeffESdS}
V_\mathrm{eff}(R)=M/R+R^2/2R_0^2
\end{equation}
with total energy $E_\mathrm{tot}=(1-E^2)/2$. This is easily seen
from writing the equation in the form $\frac{1}{2}\dot{R}^2 +
V_\mathrm{eff} (R) = E_\mathrm{tot}$. The potential is displayed
in Fig.~\ref{fig: Veff} for $M=0$ and $M=M_N/10$.
It has a minimum at $R=R_*=M^{1/3}R_0^{2/3}$. A test body has two
turning points given by the roots of the equation $f(R)=E^2$. The
larger turning radius, $R_i$, will be our comoving radial
coordinate. For the maximal energy $E_{\mathrm{tot}}=1/2$ the
turning radii are the two horizons, $R=R_+$ and $R=R_{++}$. Hence,
for the construction of coordinates in proper-time gauge we look
at geodesic motion of test bodies with initial positions $R_* <
R_i < R_{++}$ and zero initial velocity. The synchronization
condition that each geodesic passes through its turning point at
$T=\tau=0$ completes the specification of the proper-time
coordinates $(\tau,R_i)$. In terms of the new coordinates, the
energy parameter is specified simply as $E=\sqrt{f(R_i)}$.

Integration of the radial geodesic equation yields
\begin{equation}\label{ellipticIntegral}
\tau (R,R_i) = -m \int^{R_i}_R \frac{dy}{\sqrt{f(y) - f(R_i)}} = -m R_0 \int^1_\rho d\xi \sqrt{\frac{\xi}{P(\xi)}},
\end{equation}
with the dimensionless quantities $\xi=y/R_i$, $\rho=R/R_i\leq 1$,
$\beta=54 M M_N^2/R_i^3$ and
\begin{equation}
P(\xi) = -\xi^3 + (\beta+1)\xi - \beta = (1-\xi)(\xi-\xi_+)(\xi-\xi_-),
\end{equation}
\begin{equation}
\xi_{\pm} = -\frac{1}{2}\pm\Delta, \quad \Delta = \sqrt{\frac{1}{4}+\beta}.
\end{equation}
In terms of special functions the above integral becomes
\begin{equation} \label{elliptic}
\tau = -m R_0 \sqrt{2/\Delta} \left[(1-\xi_-)\Pi (\mu, \frac{\xi_+ -1}{2\Delta},r) + \xi_- F(\mu,r)\right],
\end{equation}
where
\begin{equation}
\mu = \arcsin \sqrt{\frac{ 2\Delta (1-\rho) }{ (1-\xi_+)(\rho -\xi_-)}}, \quad r = \sqrt{\frac{(1-\xi_+)(-\xi_-)}{2\Delta}}.
\end{equation}
$F$ and $\Pi$ are the elliptic functions of the first and third
kind, respectively \cite[3.167.15]{GradshteynRyzhik}.
Equation~\eqref{elliptic} is an implicit definition of
$R(\tau,R_i)$. This relation allows us to write the line element
in terms of proper-time $\tau$ and comoving spatial coordinate
$R_i$ as
\begin{equation}
ds^2 = d\tau^2 + \frac{(\partial R/\partial R_i)^2}{f(R_i)}dR_i^2 + R(\tau,R_i)^2d\Omega^2.
\end{equation}

\begin{figure}
\centering
(a)\raisebox{-180.0pt}{\includegraphics[scale=0.9]{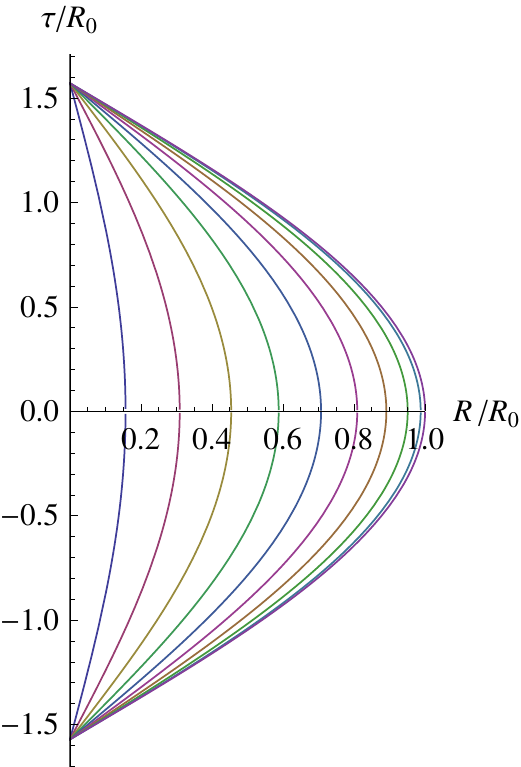}}
(b)\raisebox{-190.5pt}{\includegraphics[scale=0.9]{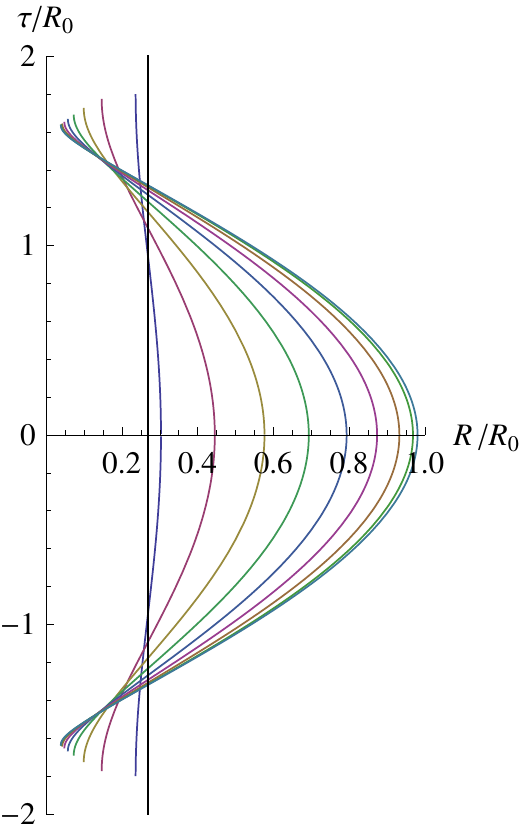}}
\caption{\footnotesize $R_i = const$ geodesics in (a) Euclidean
de~Sitter space and (b) Euclidean Schwarzschild-de~Sitter
($M=M_N/10$) space. The outermost line is the $R_i = R_{++}$
geodesic. (a) All curves converge at $\tau = \pm \pi R_0/2$. (b)
The vertical line marks $R=R_*\equiv M^{1/3}R_0^{2/3}$, the
minimum of the effective potential. Geodesics, which at $T=0$ are
located to the right of the line start intersecting each other
when entering the region $R<R_*$, forming caustics. There are no
caustics in the region $R_* < R < R_{++}$.} \label{fig:
ESdSgeodesics}
\end{figure}

When $M=0$, the expression for $R(\tau,R_i) = R_i \cos(\tau/R_0)$
is known explicitly. When $M\ne0$, the values of $R(\tau,R_i)$ can
be obtained by numerically solving the implicit
equation~\eqref{elliptic}. Both cases are shown for comparison
in Fig.~\ref{fig: ESdSgeodesics}. Note that the radial geodesics
in EdS space, Fig.~\ref{fig: ESdSgeodesics}a, do not intersect
except for extreme values of $\tau$, but those in ESdS space do,
Fig.~\ref{fig: ESdSgeodesics}b, that is, they form caustics. The
implications of these caustics are discussed in
section~\ref{causticsESdS}.

\section{Caustic formation in an interior matter solution}\label{sec:interior}

We show here that gluing the vacuum ESdS solution to an interior
solution of matter of constant density does not significantly
improve on the situation found in vacuo with respect to the
formation of caustics in a set of Gaussian normal coordinates. The
matter distribution we are interested in is the Wick-rotated
version of a simple model of a spherically symmetric relativistic
star in the presence of a positive cosmological constant. The
Euclidean stress-energy tensor can be taken to be the one of a
perfect fluid with a uniform density $\rho$,
\begin{equation}
T^{\mu\nu} = -\left(p(R)+\rho\right)u^\mu u^\nu + p(R)g^{\mu\nu}.
\end{equation}
Solving the Euclidean Einstein equations one finds that the line
element inside the star written in Schwarzschild coordinates is~\cite{Stuchlik}%
    \footnote{Some intermediate results in this reference contain
    typographical errors. However, we have verified that the formulas
    relevant for this work are correct.}
\begin{equation}\label{intmet}
ds^2 = \left(A\ Y(R_S) - B\ Y(R)\right)^2dT^2 +
\frac{dR^2}{Y(R)^2} + R^2 d\Omega^2,
\end{equation}
where we denote the location of the surface of the star by $R_S >
R_{+}$ (or $R_S > R_*$, as we restrict to later on) and define
\begin{equation} \label{ConstForInterior}
Y(R)=\sqrt{1-\frac{8\pi\rho + \Lambda}{3}R^2},~~ A=\frac{9M}{6M
+ \Lambda R_S^3},~~ B = \frac{3M - \Lambda R_S^3}{6M + \Lambda
R_S^3},~~ \rho = \frac{3M}{4\pi R_S^3}.
\end{equation}
Setting $R=R_S$ in (\ref{intmet}), we see that the interior matter
solution is matched continuously to the exterior ESdS vacuum
region. This implies that the correct matching condition for the
radial timelike geodesics is that the first derivatives match at
the surface of the star.

For the case $R_S=R_*=M^{1/3}R_0^{2/3}$ we can determine the
extension of radial geodesics to the interior region explicitly.
This is depicted in Fig.~\ref{fig: eps_one_tenth_interior}a.
\begin{figure}
\newsavebox\Abox%
\newsavebox\Bbox%
\savebox\Abox{\includegraphics[scale=0.9]{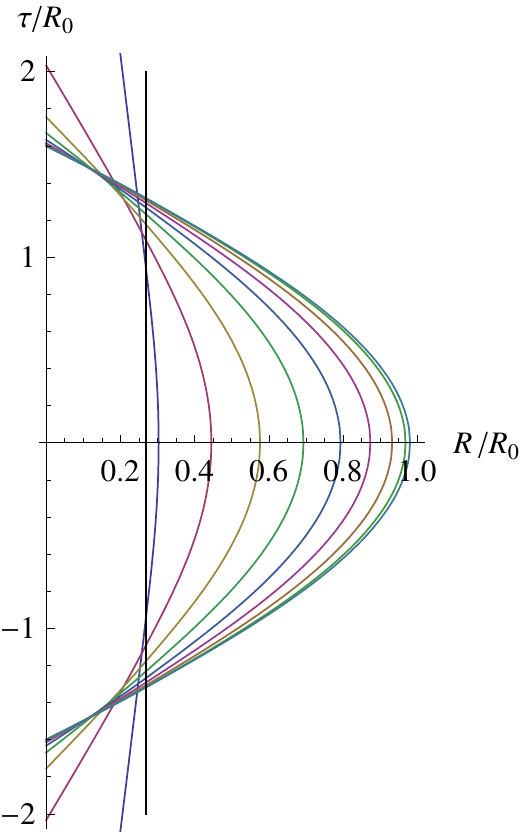}}%
\savebox\Bbox{\includegraphics[scale=0.9]{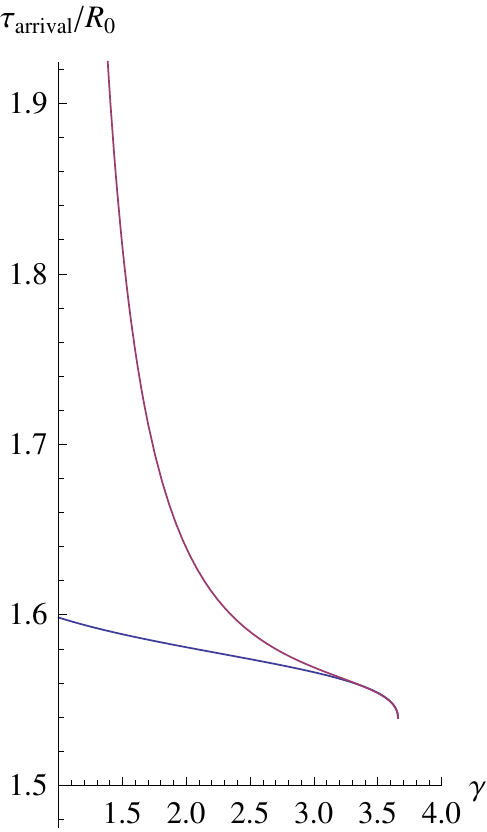}}%
\centering
(a)~\raisebox{1.5ex-\ht\Abox}{\usebox{\Abox}}
(b)~~\raisebox{1.5ex-\ht\Bbox}{\usebox{\Bbox}}
\caption{\footnotesize Gluing an interior uniform matter
distribution to an external Euclidean Schwarzschild-de Sitter
solution for $M = M_N/10$. (a) $R_i=const$ geodesics for the
special case of gluing at the radius $R=R_*$ (vertical line). The
curves intersect inside the matter region, illustrating the
breakdown of the Gaussian normal coordinates. The outermost line
is the $R_i = R_{++}$ geodesic. (b) For general gluing radius:
proper time at $R=0$ (arrival time) of the two radial geodesics,
$R=R_{++}$ (lower curve) and $R=\gamma R_*$ (upper curve), as
functions of $\gamma=R_S/R_*$, showing that adjusting the gluing
condition does not prevent the breakdown of the Gaussian
coordinates in the interior region.} \label{fig:
eps_one_tenth_interior}
\end{figure}
The geodesics intersect inside of the matter region and hence the
formation of caustics persists.

This result is not specific to the choice $R_S = R_*$. In order to
avoid a lengthy analysis we set up a simple criterion for the
intersection of geodesics. We release two test bodies, one from
$R_i = R_{++}$ and the other from $R_i = R_S$, with zero initial
velocity and compare the proper time they take to arrive at $R=0$,
which is the centre of the star. If the test body starting at the
surface takes longer, then there must be an intersection. The
arrival time for the test body starting from $R_i = R_{++}$ is
given by
\begin{equation} \label{arrivaltime}
\tau (R_i=R_{++}) = \sigma \arcsin \left(R_S/\sigma\right) +
\int_{R_S}^{R_{++}} \frac{dy}{\sqrt{f(y)}},
\end{equation}
with $1/\sigma^2 = \left(1+2/\gamma^3\right)/(27M_N^2)$ and
$\gamma = R_S/R_*$. The first term in~\eqref{arrivaltime}
constitutes the contribution of the matter region and the second
constitutes that of the vacuum region. The integral in the second
term is the same as that evaluated in~\eqref{elliptic}.

For the test body starting with zero initial velocity at the
star's surface, $R=R_S$, we have
\begin{equation}
\tau (R_i=R_S) = \int_0^{R_S}
\frac{dy}{Y(y)\sqrt{1-\frac{f(R_S)}{(AY(R_S)-BY(y))^2}}},
\end{equation}
where the constants $A, B, f(R_S)$ and $Y(R_S)$ all depend on
$\gamma$. The resulting arrival times are plotted in
Fig.~\ref{fig: eps_one_tenth_interior}b, where the upper curve
represents the $R_i = R_S$ case and was found by numerical
integration. The arrival times are equal only when the
trajectories coincide, i.e., when the star surface reaches the
cosmological horizon. Therefore, we conclude that the Gaussian
normal coordinates are not well defined in the interior region for
any choice of radius and density of the mass distribution. We have
not investigated whether fine-tuning of the internal structure of
the matter model (e.g. by considering inhomogeneous or anisotropic
fluids) could help in preventing the occurrence of caustics.
However, such a possibility appears unlikely to us.

\bibliography{esdsbib}{}
\bibliographystyle{unsrt}
\end{document}